**CHAPTER 10**

# Beyond Diversity: Computing for Inclusive Software

**Kezia Devathasan, Nowshin Nawar Arony, Daniela Damian**

This chapter presents, from our research on inclusive software within the context of a diversity and inclusion–based STEM program at the University of Victoria, INSPIRE: STEM for Social Impact (hereafter Inspire).[1] In a society with an ever-increasing reliance on technology, we often neglect the fact that software development processes and practices unintentionally marginalize certain groups of end users. While the Inspire program and its first iteration in 2022 are described in detail in [PAPER-Arony-Education], here we describe our insights from an analysis of the development processes and practices used by the teams. We found that empathy-based requirements gathering techniques and certain influences on the software development teams' motivation levels impact the teams' ability to build inclusive software. This chapter begins with an explanation of the Inspire program and a discussion on what the term "inclusive software" actually means in our context before highlighting useful practices for designing inclusive software.

## INSPIRE: STEM for Social Impact

The Inspire program at the University of Victoria was designed to empower underrepresented individuals in STEM with the ambition of increasing the retention of minority groups in STEM fields. The program consists of assigning enrolled students (both undergraduate and graduate) to a four-month-long internship on a community-led project based on students' expressed interest. Each project required students to solve a local community problem using their various STEM skills. As each project was closely tied to a real community partner, success in these undertakings yielded observable impacts for real clients, thus empowering our students. For example, we had a team of students create a website for the Victoria Brain Injury Society (VBIS) that was navigable by patients with acquired brain injuries. This allowed many patients to regain a sense of independence and autonomy over their own recovery program. We also had a team of students create a platform to help shelters coordinate available beds to assist women fleeing domestic violence or facing homelessness. As we had allowed students to choose which project they would work on, they were passionate about the cause of their projects, which has been shown to boost motivation levels [8].

---

[1] inspireuvic.org

Each team of four to five individuals were diverse in terms of academic background, and other salient characteristics, although it is important to note that students were not assigned to teams based on diversity. Among the 24 students in the program, 10 identified as female, 12 identified as male, and 2 preferred not to disclose their gender. Nineteen students came from the undergraduate level and five from the graduate level. The students were also diverse in terms of (1) academic background – they came from computer science, software engineering, electrical engineering, mechanical engineering, biomedical engineering, physics, chemistry, and business – and (2) ethnicity: our students were South Asian, East Asian, Black, Arab, Hispanic, Indigenous, and White. The teams went through several phases over the course of the four-month-long project. These phases followed the suggestions of enterprise design thinking (DT), an empathy-based approach to gathering and eliciting requirements from end users for creative projects such that end users have a voice in how a product is designed for them [2].

The students were asked to submit a weekly individual reflection in addition to a weekly team reflection. By analyzing these reflections, we were able to learn about the challenges faced by software developers in making inclusive software. Previous research already suggests that software developers find considering the needs of their diverse end users challenging [10]. Thus, we look deeper into this problem and discuss techniques that are effective in encouraging inclusive software development.

# Research Methods

The student reflections as described previously were our main source of qualitative data, in addition to several interview and focus group sessions conducted with the Inspire students. We analyzed approximately 400 reflections and 20 interview and focus group sessions using Braun and Clarke's [1] six-step thematic analysis process, which is a widely used method to identify, analyze, and report patterns or themes that represent important information from the data in relation to the research question. The steps of this method are (1) familiarizing with the data, (2) generating initial codes, (3) searching for themes, (4) reviewing potential themes, (5) defining and naming themes, and (6) producing the report [1]. We analyzed the data for evidence of challenges designing inclusive software and also looked for how students responded to the practices outlined in the following in terms of their effectiveness in helping create inclusive software.

## Inclusive Software

The term *"inclusive software"* seems intuitive upon first glance. We aim to create software that meets the needs of all potential end users of the software product. However, upon further thought, an overwhelming problem becomes apparent. The current state of knowledge defines inclusive software as software that is completely usable by *any* end user for any task [11]. To understand the magnitude of this problem, we can take a look at a widely used application such as Google Chrome. Can such an application realistically meet the needs of every possible end user? Which accessibility or usability issues should be prioritized during the development process? How should these issues be prioritized?

A problem with such a large scope is understandably difficult to tackle. As a result, modern software developments frequently fail to meet the needs of potentially diverse end users [5]. Unfortunately, this renders many software products difficult or unusable by certain user groups [5]. The idea of inclusive software itself brings an entirely new lens into the software development cycle, which traditionally consists of technical steps such as requirements defining, solution building, and testing [6]. Though effective, these traditional methods typically neglect human-centric issues. Previous work by Khalajzadeh et al. categorizes these human-centric issues into eight groups: Inclusiveness, Privacy and Security, Compatibility, Location and Language, Preference, Satisfaction, Emotional Aspects, and Accessibility. This work thus defines human-centric problems as *"the problems that diverse end users face when using a software system, due to the lack of proper consideration of their specific characteristics, limitations and abilities."* For now, we consider inclusive software as software that considers all of these human-centric issues, and we explore practices that help developers do so.

# Repairing the Digital Divide

Before delving into the practices we can employ to create more inclusive software, it is important to recognize why the concept is so important in the first place. A movement toward creating more inclusive software can help ensure that a broader range of end users can use a software product. While this may include those with disabilities, it also should consider those of different backgrounds, cultures, genders, ages, languages, etc. Furthermore, when a software product can be used by more end users, the associated organization benefits as the chances of successful adoption increase.

Software products that contain biases discriminate against certain groups of end users, both subtly (perhaps in an underlying algorithm) and explicitly (a feature of the user interface). These biases have the potential to affect the health of an individual, community, and even entire organizations [12]. A popular example in the software engineering community that highlights the detriment of non-inclusive software is gender. Previous research suggests that many software applications, everything from programming environments to educational platforms, favor males [12]. Biases in such software thus have the potential to deter other genders from this technology. Gender, however, is only one dimension of discrimination. Truly inclusive software must mitigate such discrimination across multiple dimensions.

Software that fails to be inclusive or unethical software has an incredible impact on our world. With software dictating many facets of our lives [13] that we often don't notice, an unintentional bias may be woven into our everyday lives. However, the biases seen in increasingly popular software developments such as machine learning algorithms likely come from those creating the software products [13]. For example, in 2015, Flickr's image recognition feature was flagged for yielding racist results, as it was unable to accurately identify those of African descent [13]. This raises questions on whether or not the training set for this algorithm included enough samples of diverse races in the first place. With these considerations in mind, it becomes apparent that much of a team's biases affect the inclusiveness of the software that they produce. This leaves us with the question: *despite inevitable human biases, what software development practices help teams create inclusive software?*

# Design Thinking for Inclusive Software

All five projects in the Inspire program had a very unique subset of end users, each containing different dimensions of diversity that the teams had to consider. In the Inspire program, we relied on the expertise of each community partner to help us decide on how inclusive the software produced by each student team was. For example, the team that created a website for patients with acquired brain injury had to consider many needs of various patients, as brain injuries are each unique in terms of the cognitive deficits they impose. The community partner at the Victoria Brain Injury Society was pleasantly surprised at how accommodating the features of the website were. Another Inspire project resulted in a system to track visitation patterns to a local nature sanctuary in order to protect endangered species. The community partner for this project was again delighted with how our students were able to solve this problem in a manner that worked for all potential site visitors and respected all privacy regulations and concerns. Our third project required Inspire students to come up with creative requirements elicitation methods to engage youth in discussions surrounding climate change and climate change anxiety. This was especially challenging as youth are particularly vulnerable to negative emotions surrounding climate change and can also be more challenging to engage for sustained periods of time. The fourth project in Inspire was especially concerned with gathering requirements in a delicate and respectful way, as this team was engaging with women+ individuals fleeing domestic violence and/or experiencing homelessness. Students on this project had to be prepared to deal with different levels of severity of the client's personal situation and tread carefully to avoid offensive or triggering language. The fifth Inspire project dealt with creating a platform to show climate risk zones in our local region (i.e., highlighting areas prone to flooding, landslides, and other natural threats). In this case, the potential end users for this project were anyone in our local community, and the team was tasked with being inclusive of varying levels of technological literacy.

# IBM Design Thinking

Success in the Inspire program can be attributed to the software design methodologies we taught our students to employ. IBM's version of design thinking (DT) and resources on the process were provided and taught to our students. As previously mentioned, software development strategies such as Agile are very effective but often neglect the needs of the end users. This is where design thinking (DT) is particularly effective.

DT is a development methodology that involves prioritizing the feelings and wants of the intended end user *during* the requirements gathering process as opposed to treating human-centric requirements as an afterthought [7]. Typically when software is developed, the main features are implemented, and accessibility options are then added into the interface. However, in considering users' needs during requirements elicitation and implementation, there is an improved emphasis on satisfying user needs [7].

The phases of DT, also employed by the Inspire program's students, are as follows [7]:

- **Understand**: *Focus on gaining empathy for end users' pain points.*
- **Explore**: *Generate solution ideas while avoiding overly simplistic solutions.*
- **Prototype**: *Iterative series of mockups to visualize ideas from the Explore phase.*
- **Evaluate**: *Gather feedback on generated prototypes.*

DT is unique as it is a human-centered approach to software development that places a heavy emphasis on empathy. For the purposes of DT methodologies, empathy can be defined as the ability for the software development team to truly understand the needs and wants of the product's end users. Due to its human-centric focus, DT relies on a team's potential multidisciplinary talents. For example, in the Inspire program, the ability to program a website was only a small subset of the total skill required to accomplish a project. Students needed to be able to establish good communication with their community partner, negotiate project scope in a professional manner, and learn critical requirements gathering techniques such as conducting focus groups and interviews when dealing with highly sensitive topics around vulnerable persons.

## Practices of Design Thinking

In this section of the chapter, we discuss several requirements gathering techniques that are unique to DT and that were used in the Inspire program. Each of these techniques introduces a new way to empathize and understand potential end users, thus giving software teams a better perspective on the needs of their clients.

## Empathy Mapping

The development of effective communication skills was critical to the results of highly inclusive software that we observed. DT has several requirements elicitation techniques; a popular practice is empathy mapping. When creating an empathy map, the development team discusses pain points with a potential end user of a software product. Utterances the end user makes during such a session are then categorized into feelings, thoughts, actions, and statements. An empathy map helps the development team identify how an individual is currently navigating their pain points and helps yield creative ideas for the prototyping phase. A participant witnessing the development team categorizing their statements also helps curate a space where they feel cared for and heard, encouraging the participant to share their thoughts more openly and honestly.

Students in the Inspire program used empathy mapping frequently for their projects and were able to gain a deeper understanding into how their clients were navigating their pain points. Figure 10-1 shows the students using an empathy map to identify the common data points from the collected data. One student said empathy mapping really helped their team *"figure out what we wanted to get out of our interviews and helped us figure out how we could craft those questions in a respectful way."*

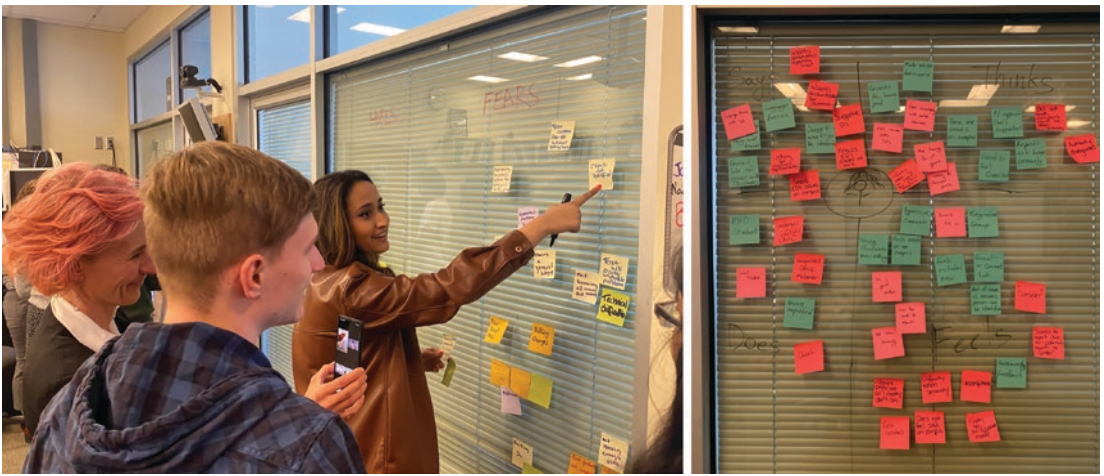

***Figure 10-1.*** *Inspire students presenting on what they have learned about their end user's needs after empathy mapping*

## Hopes and Fears

Another common practice is called *hopes and fears*. In this exercise, the development team makes a list of all their greatest aspirations and worries for the designated project. When this exercise was done in the Inspire program, an overwhelming amount of both hopes and fears from students were related to the satisfaction of the end users. For example *"I hope that we can build a product that actually makes a difference for people facing homelessness."* This demonstrates that a lot of strong emotions toward the project are centered around satisfying end user needs. Identifying a team's greatest hopes and fears for the project may be helpful in deciding which requirements elicitation methods to use and which to avoid. For example, if a software is to be designed for vulnerable groups such as individuals facing homelessness, and a team is nervous about interviews with such persons, then a greater emphasis can be placed on fine-tuning an interview structure or finding another requirements gathering technique altogether. This is critical in creating an inclusive requirements gathering experience, so that clients feel they are in a safe space to share their true needs and wants for an ultimately inclusive end product.

## Journey Mapping

An important technique in DT that helps developers empathize with how end users interact with software is journey mapping. Simply put, this is a visualization of what a user experiences when they interact with the software to accomplish a goal. This timeline of steps is then embellished with things like the user's thoughts or emotions at specific moments in this journey. In the Inspire program, our students learned how overwhelming it could be for vulnerable individuals to use the software they were currently using through journey mapping. In Figure 10-2 the students can be seen using a journey map. This catalyzed discussions around simplifying user interfaces and having clearer instructions and even subtler discussions surrounding simplistic fonts and colors. Creating inclusive software is more than just creating the software; other elements such as customer communication and even marketing must also be inclusive, and journey mapping is a way to achieve this [4].

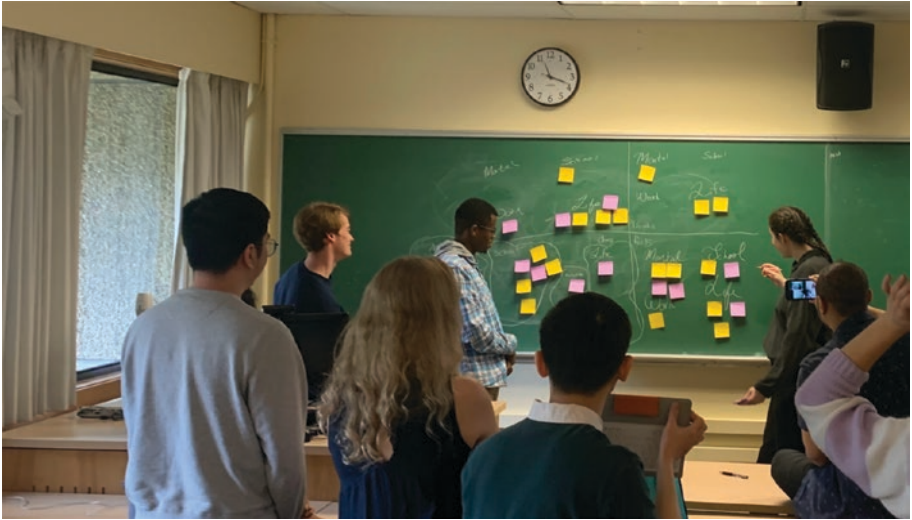

*Figure 10-2.  Inspire students explaining insights from journey mapping*

## Playbacks

Playbacks are informal presentations that are done at regular intervals to update sponsors, clients, and other project members on the progress of the project. Within the Inspire program, these playbacks were a safe space to receive feedback from other students on angles the team might have missed about their end users and from their community partners. These sessions were also conducted to help keep Inspire teams on track in terms of scope and feasibility of their prototypes. These sessions were particularly insightful as our students shared what they had learned from their clients with the Inspire teams. It was incredible to see how their assumptions about their end user needs and wants had been disproven and how the practices in DT had led them to uncover the true depth of many of these community problems.

## Revisiting Previous Practices

In addition to these unique DT practices, revisiting some traditional techniques such as interviews and focus groups with a design thinking lens is just as important. In Part 2, [Chapter 6, "Elicitation Revisited for More Inclusive Requirements Engineering"], Tizard et al. acknowledge that diverse end users may not respond as well to traditional elicitation methods. With respect to DT, while traditional methods are still employed, they take on much more forethought than the typical preparation for an interview or focus groups. Figure 10-3 shows the students at a school event to gather requirements from teens for their project.

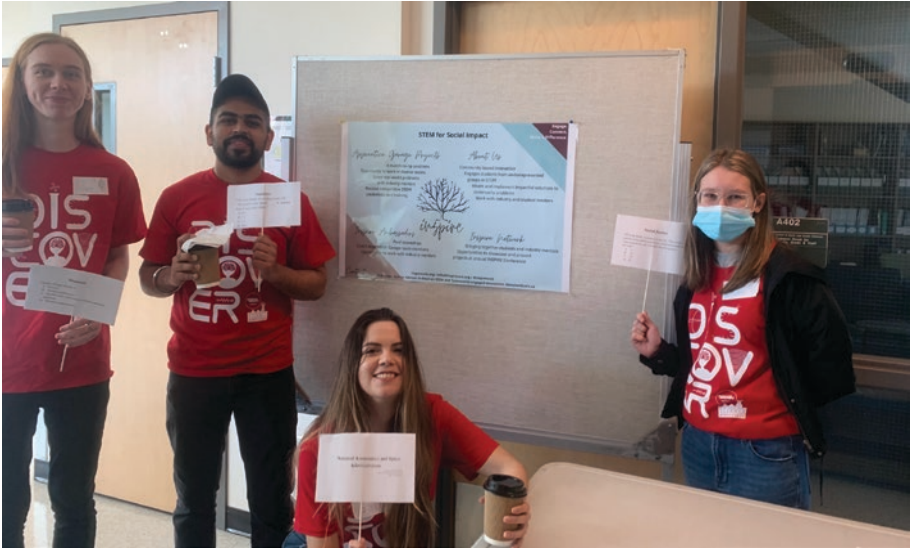

*Figure 10-3.  Students in the Inspire program experienced much success revisiting traditional elicitation techniques with end users in mind. The ClimAct team (pictured) was especially concerned with engaging youth and ended up gathering requirements by gamifying the interview process.*

Our students in the Inspire program spent ample time focusing on the language surrounding their focus group and interview questions and received project-specific training on how to interact with their clients. For example, interviews conducted with patients with acquired brain injuries were much different than interviews conducted with neurotypical individuals. Similarly, gender-neutral language was imperative in sessions involving women+ individuals facing homelessness. Students in the Inspire program initially described having to interview vulnerable participants as *"overwhelming"* and *"intimidating,"* but found that the time spent finessing the interview dialogue and environment helped them *"understand their participants better* and *"really get our end users to open up"* to them.

## Inclusive Teams, Inclusive Software

The point of inclusive software is to design a product that caters to diversity within an intended end user base. In one of our Inspire projects, for example, the website designed for patients with acquired brain injuries was not designed with the average, neurotypical person in mind. There can be an incredible amount of diversity even within a group of specific end users in mind, and creating inclusive software means acquiring an understanding of their specific characteristics and needs.

In Inspire we found that the true significance of the inclusive design thinking practices outlined previously is actually found within the software development team itself, not the end users.

While inclusive requirements elicitation is indeed critical in creating inclusive software, we discovered that a prerequisite to inclusive software is inclusion within the software development team. For example, the hopes and fears exercise catalyzed discussions within the team of each other's strongest emotions regarding the project, facilitating a sense of empathy within team members. While we recognized and focused on the diversity of our student teams, it is the inclusion within these teams that made diversity significant. In a heterogeneous group, the many advantages of diversity such as greater possibility to understand and represent the final user needs [3] and generation of more unique ideas when compared with homogeneous teams [9] cannot be leveraged without inclusion.

Our observations suggest that the empathy-based design thinking processes facilitated a sense of empathy within team members and a greater ability to understand different perspectives. In one student's words, *"…We have really developed so much empathy for each other in the group; understanding each others strengths, weaknesses, and emotions has been crucial so far."* Such empathy for each other allowed teams to ultimately be empathetic to their end users. An Inspire student mentioned that the empathy they had for their team members "*really helped highlight how important the design thinking practices are. It even helped with creating our interview and focus group questions with an empathetic lens to make our clients more comfortable too."* Another student mentioned, *"In using the design thinking methods we were taught in Inspire, I could see that there was more than one user group; there were actually three main categories of users based on their life stages and personal journeys. It really made me think about whether or not the platform we were building would be viable or useful for each of these user groups."* This suggests that students were developing a heightened sense of empathy within their teams throughout the design thinking process, which they were able to then apply to requirements elicitation with their end users.

## Empathy and Team Morale

The development of empathy within teams that was facilitated by DT techniques assisted the teams in empathizing with their end users and thus helped the teams create inclusive software. Creating inclusive software this way can be an arduous task; many iterations and pivots (i.e., changes in project directions) occur, and team frustration is rather inevitable. A student mentioned that their team experienced many pivots and expressed that at some points along the project they felt *"defeated."*

Aside from the direct relationship to creating inclusive software, the sense of empathy that the students developed also dramatically helped with team morale. It was a mitigating factor when teams experienced turbulent times. The same student who described feeling *"defeated"* subsequently said, *"I would not have been able to overcome this without my team. They are very empathetic and can tell if I'm having a bad day and need to take on a lighter workload for the time being."* Such a sense of empathy thus increased team morale overall, which also improved motivation. As the students were able to work well together in terms of team dynamic, they were able to fully focus on the parts of the project that excited them the most, thus boosting their overall motivation [8].

## A Step Toward Inclusive Software

Recall that inclusive software largely encompasses an end user's emotions [5]; thus, DT techniques such as empathy and journey mapping can contribute dramatically to making the entire software development process more user focused in order to create software that is more inclusive. The first year of the Inspire program was a stepping stone in learning how software teams approach inclusive software development. Applying the DT practices in real-world scenarios for diverse end users allowed the students to learn and empathize. The students recognized their personal biases and emphasized the users' needs throughout the product development. While our teams superseded our initial expectations and maximized the potential of DT practices, much more work in the area of inclusive software has to be done to truly understand how we can curate an inclusive digital environment.

## Conclusion and Takeaways

Overall, an analysis of performance in the Inspire program suggests that empathy-based approaches are in fact effective at instilling empathy in software development teams. Moreover, we see *why* this is so important: it allows for the creation of software that caters to a diverse set of end users. While fully introducing DT methodologies into existing software development pipelines may not be feasible for all organizations, we leave the reader with the following points of consideration:

- A sense of empathy within the software development team must be fostered before inclusive software can be successfully developed.

- Some design thinking practices can still be implemented to facilitate this sense of empathy, even if the full design thinking methodology is not adopted.

- A developed sense of empathy is critical to high team morale, especially in diverse software engineering teams.

- Creating inclusive software via these empathy-based techniques is ultimately required to repair the digital divide observed today.

## Bibliography

[1] Virginia Braun and Victoria Clarke. Using thematic analysis in psychology. *Qualitative Research in Psychology*, 3(2):77–101, January 2006. Publisher: Routledge eprint: www.tandfonline.com/doi/pdf/10.1191/1478088706qp063oa.

[2] Sasha Costanza-Chock. *Design Justice: Community-Led Practices to Build the Worlds We Need*. The MIT Press, 2020.

[3] John Grundy, Ingo Mueller, Anuradha Madugalla, Hourieh Khalajzadeh, Humphrey O. Obie, Jennifer McIntosh, and Tanjila Kanij. Addressing the influence of end user human aspects on software engineering. In *International Conference on Evaluation of Novel Approaches to Software Engineering*, 241–264, Springer, 2021.

[4] Tharon Howard. Journey mapping: A brief overview. *Communication Design Quarterly Review*, 2(3):10–13, 2014.

[5] Hourieh Khalajzadeh, Mojtaba Shahin, Humphrey O. Obie, and John Grundy. How are diverse end-user human-centric issues discussed on github? Preprint at *arXiv:2201.05927*, 2022.

[6] Yu Beng Leau, Wooi Khong Loo, Wai Yip Tham, and Soo Fun Tan. Software development life cycle agile vs traditional approaches. In *International Conference on Information and Network Technology*, volume 37, 162–167, 2012.

[7] Percival Lucena, Alan Braz, Adilson Chicoria, and Leonardo Tizzei. IBM design thinking software development framework. In *Brazilian Workshop on Agile Methods*, 98–109, Springer, 2017.

[8] Michael Miles, David Melton, Michael Ridges, and Charles Harrell. The benefits of experiential learning in manufacturing education. *Journal of Engineering Technology*, 22(1):24, 2005.

[9] Charlan J. Nemeth. Differential contributions of majority and minority influence. *Psychological Review*, 93(1):23, 1986.

[10] John Noll, Sarah Beecham, Abdur Razzak, Bob Richardson, Ann Barcomb, and Ita Richardson. Motivation and autonomy in global software development. In *International Workshop on Global Sourcing of Information Technology and Business Processes*, 19–38, Springer, 2017.

[11] Anthony Savidis and Constantine Stephanidis. Inclusive development: Software engineering requirements for universally accessible interactions. *Interacting with Computers*, 18(1):71–116, 2006.

[12] Mihaela Vorvoreanu, Lingyi Zhang, Yun-Han Huang, Claudia Hilderbrand, Zoe Steine-Hanson, and Margaret Burnett. From gender biases to gender-inclusive design: An empirical investigation. In *Proceedings of the 2019 CHI Conference on Human Factors in Computing Systems*, 1–14, 2019.

[13] Adrienne Yapo and Joseph Weiss. Ethical implications of bias in machine learning. 2018.